\documentclass[prl,showpacs,nofootinbib,twocolumn]{revtex4}
\usepackage{graphicx,color}% Include figure files

\begin{document}

\title{The Clean Way to Identify a Scalar Glueball}
\author{ Wei Wang$^{a}$ \footnote{Email: wwang@ihep.ac.cn},
Yue-Long Shen$^{b}$ \footnote{Email: shenyl@ihep.ac.cn}
 and Cai-Dian L\"u$^{a}$ \footnote{Email: lucd@ihep.ac.cn}}
\affiliation{
 $^a$ Institute of High Energy Physics  and Theoretical Physics Center for Science Facilities,
 Chinese Academy
of Sciences, Beijing 100049, People's Republic of China \\
 \it $^b$ College of Information Science and Engineering,
Ocean University of China, Qingdao, Shandong 266100, People's Republic of China%\\
% \it $^d$ Theoretical Physics Center for Science Facilities,  Chinese Academy
%of Sciences, Beijing 100049, People's Republic of China
}

\begin{abstract}

The existence of a glueball has been a tough work for many years
study.  Utilizing the well developed  QCD theory for $B$ meson
decays, we propose a new way to identify whether  a scalar glueball
existed or not. In the presence of mixing between glueballs and
ordinary scalar mesons, we   explore the possibility to extract the
mixing parameters from semileptonic $B$ decays and nonleptonic $B$
decays. We also point out a clean way to identify a glueball through
$B_c$ decays.
\end{abstract}

\pacs{13.20.He,14.40.Cs} \maketitle

\section{Introduction}

Quark model has achieved a great success to describe hadronic
states, but the QCD also predicts the existence of mesons without
any valence quark, which is called glueball. The confirmation of a
glueball is one of the most important topics in hadron physics and
this subject has received extensive interests \cite{rev}. Lattice
QCD, which is almost the only method to do calculations from the
fundamental QCD, predicted that the mass of the lowest-lying scalar
glueball ($0^{++}$) is around 1.5-1.8 GeV~\cite{Bali:1993fb}.
Several different candidates have been observed in this mass region,
but there is not any solid evidence on the existence of a pure
glueball. For decades, people have tried to find a way to verify the
existence of a glueball through its decay property. The glueball is
quark flavor singlet, which should decay to $u\bar u$, $d\bar d$ and
$s\bar s$ equally. However, the claimed ``unique'' feature of quark
flavor singlet, is not unique, since the quark-antiquark state can
also be flavor SU(3) singlet. Thus, it is not a firm evidence for a
glueball. Furthermore, it is very likely that the glueball mix with
the quark-antiquark scalar  state and they together form several
physical mesons. On the theoretical side, there are large
ambiguities on the mixing mechanism~\cite{scalarmesons}. This makes
the study even more complicated.

Recently, another important direction to uncover the mysterious
structure of scalar mesons, is discussed, which   investigates the
scalar meson production property  through $B$ meson decays. In $B$
meson decays, the $O_{8g}$ operator has a large Wilson coefficient,
which could produce a number of gluons. These gluons in the final
state may have the tendency to form a glueball state, thus the
glueball production in inclusive $B$ decays has attracted some
theoretical interest~\cite{Minkowski:2004xf,He:2006qk}.  In a recent
study \cite{first}, we calculate the   transition form factors of
$B$ meson decays into a scalar glueball in the light-cone formalism.
Compared with form factors of $B$ to ordinary scalar mesons, the
$B$-to-glueball form factors have the same power in the expansion of
$1/m_B$. Taking into account the leading twist light-cone
distribution amplitude, we find that they are numerically only a
little smaller than those form factors of $B$ to ordinary scalar
mesons. It means that the production rate of glueball in B decays is
quite copious.

The scalar meson can be produced in $B$ decays by two gluon
(glueball) and also an isosinglet $q\bar q$ pair (ordinary meson).
In this paper, we will propose a method for experiments to measure
the nonzero two gluon contribution, so that to prove the existence
of a
 scalar glueball cleanly. This will require the detection of scalar
 meson
 production from $B_c$ decays.

\section{the study of mixing between glueball and quark states}

\begin{figure}
\begin{center}
\includegraphics[scale=0.5]{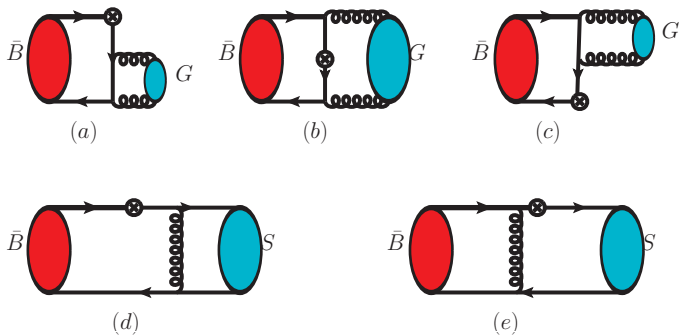}
\end{center}
\caption{Feynman diagrams of $\bar B$ decays into a scalar glueball
$G$ and an ordinary scalar meson. The $\otimes$ denotes the possible
Lorentz structure arising from the electroweak interactions. }
\label{diagram:FeynBtof0}
\end{figure}

Up to the leading Fock state, a glueball is made up of two
constituent gluons. In exclusive $B$ decays, these two gluons can be
emitted from either the heavy $b$ quark or the light quark. In the
expansion of $\alpha_s$, the lowest order Feynman diagrams for form
factors of $B$ decays into a scalar glueball are depicted in
Fig.~\ref{diagram:FeynBtof0} (a), (b) and (c). In
Fig.~\ref{diagram:FeynBtof0} (d) and (e), the light antiquark in $B$
meson and the energetic quark from the electro-weak vertex  also
form an isospin or SU(3) singlet scalar meson. Usually, people
believe that the form factor of $B$ decays to an ordinary scalar
meson is larger than that of $B$ decays into a scalar glueball. Our
recent study shows \cite{first} that the form factors of $B$ decays
into a scalar glueball is big enough for the experiments to observe
it. Compared with our previous studies~\cite{Wang:2006ria,Li:2008tk}
on the transition form factors of $B$ mesons decays into ordinary
scalar mesons (denoted as $f_0$ with the mass around 1.5 GeV), the
$B\to G$ form factors are at the same order of magnitude. The
$B$-to-glueball form factors are only a factor of two smaller than
the $B\to \pi$ form factors. These form factor results are collected
in table~\ref{form} for comparison~\footnote{  If scalar mesons
$f_0$ are identified as $\bar qq$ excited states, referred as
scenario I, the decay constants of $f_0$ are negative and so are
$B\to f_0$ form factors. In scenario II, where scalar mesons $f_0$
are identified as $ \bar qq$ ground state, the form factors are
positive.}. In fact, the main decay channel of a scalar glueball is
$\pi\pi$ or $K\bar K$. Thus a scalar glueball is much easier to
detect than the iso-singlet pseudoscalar meson such as $\eta$.
Compared with the recently measured semileptonic $B\to \eta$
decay~\cite{Aubert:2008ct}
\begin{eqnarray}
 {\cal B}(B^-\to \eta l^-\bar\nu)&=&(3.1\pm0.6\pm0.8)\times
 10^{-5},
\end{eqnarray}
the branching ratio of $B\to G l\bar\nu$~\cite{first} is comparable
with that of $B\to \eta l\bar\nu$ decay and may be observed on the
ongoing $B$ factories. It is very likely for the forthcoming Super B
factory to observe a pure glueball, if it exists.

%%%%%%%%%%%%%%%%%%%%%%%%%%%%%%%%%%%%%%%%%%%%%%%%%%%%%%%%%%%%%%%%%%%%%%%%%%%%%%
%%%%%%%%%%%%%%%%%%%%%%%%%%%%%%%%%%%%%%%%%%%%%%%%%%%%%%%%%%%%%%%%%%%%%%%%%%%%%%

%%%%%%%%%%%%%%%%%%%%%%%%%%%%%%%%%%%%%%%%%%%%%%%%%%%%%%%%%%%%%%%%%%%%%%%%%%%%%%
\begin{table}
\caption{$B$ to glueball $(G)$, $B$ to ordinary scalar meson ($f_0$)
and $B$ to pseudoscalar meson ($\pi$) transition form factors in the
PQCD approach}
 \begin{tabular}{|c|c c |}
 \hline
 \ \ \        &$F_0(0)=F_1(0)$  &$F_T(0)$    \\
 \hline
 $B\to G$\ \ \       &$0.11^{+0.02}_{-0.02}$   &$0.05^{+0.01}_{-0.01}$   \\
 \hline
 $B \to \pi$\ \      &$0.22^{+0.04}_{-0.05}$   &$0.27^{+0.05}_{-0.06}$  \\
 \hline
 $B \to f_0$ Scenario I\ \      &$-0.30_{-0.09}^{+0.08}$   &$-0.39_{-0.11}^{+0.10}$ \\
     $B \to f_0$ Scenario II   &$0.63_{-0.14}^{+0.23}$   &$0.76_{-0.17}^{+0.37}$\\
 \hline\hline
 \end{tabular}
\label{form}
\end{table}
%%%%%%%%%%%%%%%%%%%%%%%%%%%%%%%%%%%%%%%%%%%%%%%%%%%%%%%%%%%%%%%%%%%%%%%%%%%%%%

However, there is not any solid experimental evidence for a pure
glueball state up to now. The reason may be that the glueball state
can mix with the ordinary meson through the strong interactions. For
example, the Lattice QCD collaboration predicted the mass of a
scalar glueball ground state around $1.5$-1.8 GeV. It is very likely
that the glueball state mix with the ordinary quark-antiquark state
and they form several physical mesons. In this mass region, there
are three scalar mesons: $f_0(1370)$, $f_0(1500)$ and $f_0(1710)$,
which might be the potential candidates. The mixing matrix can be
set as
\begin{eqnarray}
 \left( \begin{array}{c}
 f_0(1710) \\f_0(1500)\\ f_0(1370)
 \end{array} \right) = \left( \begin{array}{ccc} a_1 & a_2
&a_3 \\b_1 &b_2 &b_3  \\c_1 &c_2 &c_3\end{array}\right)
\left(\begin{array}{c} G \\ \bar ss \\ \bar nn\end{array}
\right).\label{1}
\end{eqnarray}
For each physical scalar meson for example $f_0(1370)$, which is a
mixture of glueball and ordinary  states, the coefficients $c_1$,
$c_2$ and $c_3$ satisfy the normalization condition
\begin{equation}
\sqrt{|c_1|^2+|c_2|^2+|c_3|^2}= 1.\label{ren}
\end{equation}
A non-zero $c_1$ would be a clear evidence for the existence of a
glueball. Let us aim this to see if there is a way to settle it in
$B$ decays. The semileptonic $B\to f_0l\bar\nu$ decays receive
contributions from the $\bar nn$ component but without $\bar ss$
component (at least negligible), while the semileptonic $B_s\to f_0
l^+l^-$ channel only receive contributions from the $\bar ss$ but
without $\bar nn$ component. Both of  the decay channels can receive
gluon component contributions. Thus from eq.(\ref{ren}), we notice
that the two independent mixing parameters can be fitted from the
above two experimental measurements, in principle. For the three
kinds of $f_0$'s, we have altogether 6 experiments, but only three
real parameters in eq.(\ref{1}) to be fixed. Since the branching
fraction of $B_s\to f_0l^+l^-$ is expected to have the order of
$10^{-8}$ or even smaller, one needs to accumulate a large number of
$B$ decay events. This could be achieved on the future experiments
such as the Super B factory.

Semileptonic $B$ decays are clean but in $B\to f_0l\bar\nu$, the
neutrino is identified as missing energy and the efficiency is
limited; while the $B_s\to f_0l^+l^-$ has a small branching ratio.
In these decays, the lepton pair does not carry any SU(3) flavor and
the decay amplitudes receive less pollution from the strong
interactions. The lepton pair can also be replaced by a charmonium
state such as $J/\psi$ since $J/\psi$ does not carry any light
flavor either. $B\to J/\psi f_0$ decays may provide another ideal
probe to detect the internal structure of the scalar mesons. In
$B\to J/\psi f_0$ decay, the $\bar ss$ component will not contribute
at the leading order in $\alpha_s$. For example, the $B\to
J/\psi\phi$ decay has been set a very stringent upper
limit~\cite{:2008bt}: $ {\cal B}(B\to J/\psi\phi)<9.4\times
10^{-7}$. Thus $B\to J/\psi f_0$ decay can filter out the glueball
component and the $\bar nn$ component of a scalar meson. Meanwhile
in $B_s\to J/\psi f_0$ decay, only the $\bar ss$ and the gluon
component contributes. Moreover, the final mesons in these channels
are easy to reconstruct and these channels could have sizable
branching fractions. If we use the factorization method, decay
amplitudes are given as
\begin{eqnarray}
 A(\bar B^0\to J/\psi f_0)&=&\frac{G_F}{\sqrt 2} V_{cb}V_{cd}^* m_B^2
 f_{J/\psi} a_2 F_1^{B\to f_0}(m_{J/\psi}^2).\nonumber\\
\end{eqnarray}
The Wilson coefficient $a_2$ can be extracted from the $B\to J/\psi
K$ decays~\cite{Amsler:2008zzb}
\begin{eqnarray}
 {\cal B}(\bar B^0\to J/\psi \bar K^0)&=&(8.71\pm0.32)\times
 10^{-4}.
\end{eqnarray}
The branching ratios are roughly predicted as
\begin{eqnarray}
 {\cal B}(\bar B^0\to J/\psi f_0(\bar nn))
 %&=&\left |\frac{V_{cd}}{V_{cs}}\right|^2\left(\frac{F_1^{B\to f_0(\bar nn)}(m_{J/\psi}^2)}
 %{F_1^{B\to K}(m_{J/\psi}^2)}\right)^2\times (8.71\pm0.32)\times
% 10^{-4}\nonumber\\
 &\simeq&\left\{\begin{array}{c}
      (23 ^{+12}_{-14})
 \times 10^{-6}\;\;\;\;\;{\rm S1}\\
        (10^{+7}_{-5})
 \times 10^{-5}\;\;\;\;\; {\rm S2}
      \end{array}\right.,\\
 {\cal B}(\bar B^0\to J/\psi G)
 %&=&\left |\frac{V_{cd}}{V_{cs}}\right|^2
 %\left(\frac{F_1^{B\to G}(m_{J/\psi}^2)}{F_1^{B\to K}(m_{J/\psi}^2)}\right)^2\times(8.71\pm0.32)\times
% 10^{-4}\nonumber\\
 &\simeq& (6.2\pm2.2) \times 10^{-6},
\end{eqnarray}
where we have assumed the same $q^2$ dependence for all form factors
and $F_1^{B\to K}(0)=0.3$. The  uncertainties are from the
experimental data for ${\cal B}(B\to J/\psi K)$ and   the $B\to S$
form factors at the $q^2=0$ point.  For the $B_s$ decays, the
branching ratios are comparable with that of $B\to J/\psi K$:
\begin{eqnarray}
 {\cal B}(\bar B_s\to J/\psi f_0(\bar ss))&\simeq & \left\{\begin{array}{c}
     (6.5^{+4.0}_{-4.5})
 \times 10^{-4}\;\;\;\;\;\;\; {\rm S1}\\
        (3.5 ^{+2.3}_{-1.4})
 \times 10^{-3}\;\;\;\;\;\;\; {\rm S2}
      \end{array}\right.,\\
 {\cal B}(\bar B_s\to J/\psi G)&\simeq& (9.7\pm3.9)\times 10^{-5}.
\end{eqnarray}
Such large branching fractions offer a great opportunity to probe
structures of scalar mesons. With the available data in the future,
the mixing problem between the scalar mesons will be solvable and
the glueball component can be projected out in principle.

%%%%%%%%%%%%%%%%%%%%%%%%%%%%%%%%%%%%%%%%%%%%%%%%%%%%%%%%%%%%%%%%%%%%%%%%%%%%%%
\begin{figure}
\begin{center}
\includegraphics[scale=0.4]{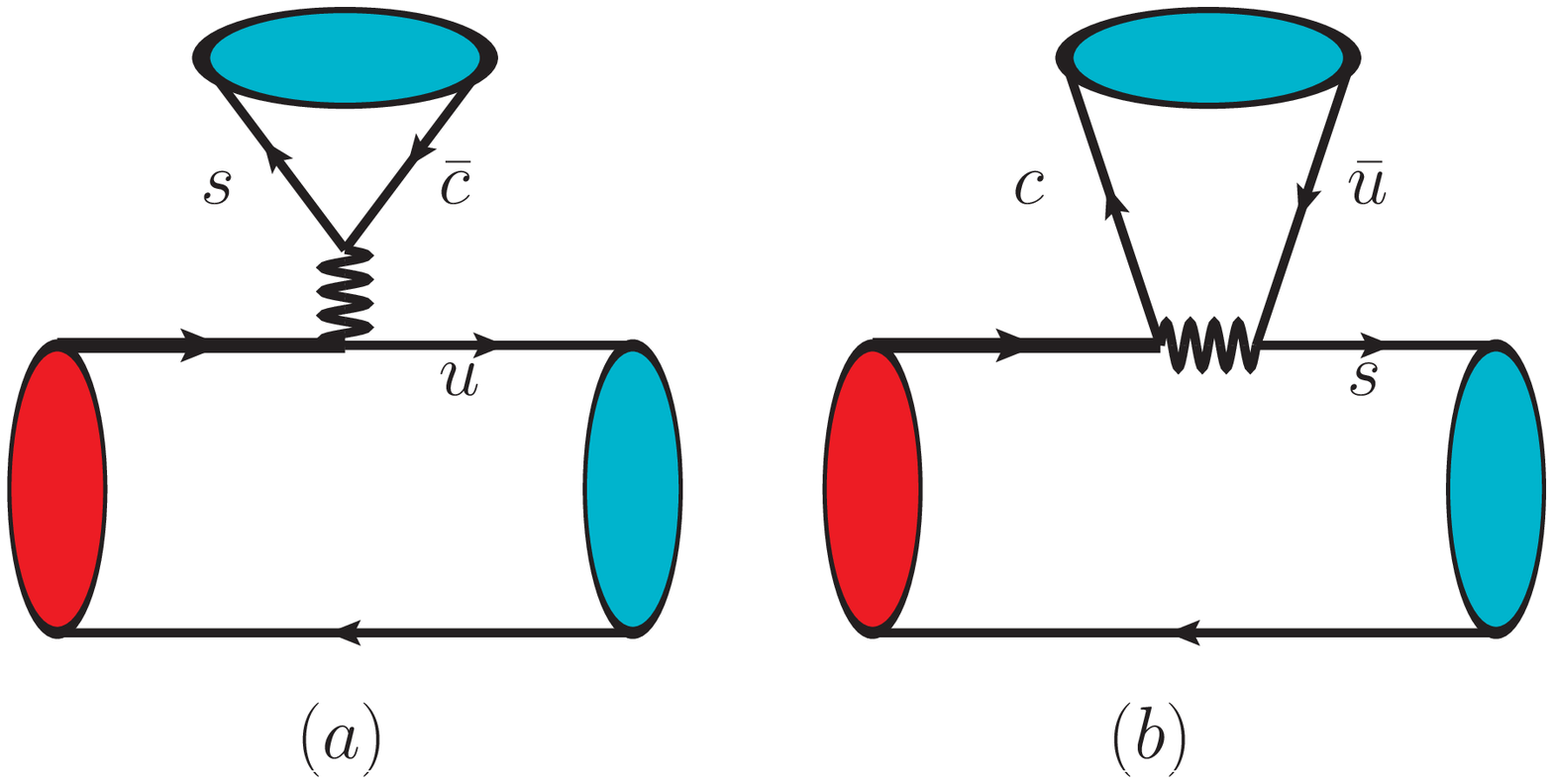}
\end{center}
\caption{Feynman diagrams of $B\to f_0D$ decays. }
\label{diagram:FeynBtof0D}
\end{figure}
%%%%%%%%%%%%%%%%%%%%%%%%%%%%%%%%%%%%%%%%%%%%%%%%%%%%%%%%%%%%%%%%%%%%%%%%%%%%%%

If the power-suppressed annihilation diagrams are neglected, the
charmful decays of $B$ meson, $B\to f_0D$, can also be used to
constrain the mixing between scalar mesons. For instance in $B^-\to
D_s^- f_0$, the $\bar nn$ and gluon component contribute but the
$\bar ss$ component does not, while in $\bar B_s\to D^0 f_0$, the
$\bar nn$ component will not contribute, as shown in
Fig.~\ref{diagram:FeynBtof0D}. Thus the mixing coefficients can also
be determined if these two channels are experimentally measured. It
is necessary to point out that this method may suffer from sizable
uncertainties of annihilation diagrams \cite{Wang:2006ria}.

To be more specific, we will discuss two mixing mechanisms in
detail. Because the decay width of $f_0(1500)$ is not compatible
with the ordinary $\bar qq$ state, Amsler and Close claimed that
$f_0(1500)$ is primarily a scalar glueball~\cite{Amsler:1995tu}. In
the subsequent studies, they extracted the mixing matrix through
fitting the data of two-body decays of scalar
mesons~\cite{Close:2000yk}:
\begin{eqnarray}
 \left( \begin{array}{c}
 f_0(1710) \\f_0(1500)\\ f_0(1370)
 \end{array} \right)
 = \left( \begin{array}{ccc} 0.36 & 0.93&0.09 \\
 -0.84 &0.35 &-0.41  \\
 0.40 &-0.07 &-0.91\end{array}\right)
\left(\begin{array}{c} G \\ \bar ss \\ \bar nn\end{array}
\right).\label{2}
\end{eqnarray}
Based on the SU(3) assumption for scalar mesons and the quenched
LQCD results, Cheng et al.~\cite{Cheng:2006hu} reanalyze all
existing experimental data and fit the mixing coefficient as
\begin{eqnarray}
 \left( \begin{array}{c}
 f_0(1710) \\f_0(1500)\\ f_0(1370)
 \end{array} \right)
 = \left( \begin{array}{ccc} 0.93 & 0.17&0.32 \\
 -0.03 &0.84 &-0.54  \\
 -0.36 &0.52 &0.78\end{array}\right)
\left(\begin{array}{c} G \\ \bar ss \\ \bar nn\end{array}
\right).\label{3}
\end{eqnarray}
 It is
found that the $f_0(1710)$ tends to be a primary glueball. This is
very different from the first matrix of mixing coefficients in
(\ref{2}). The scalar meson production rates in $B$ meson decays can
be used to distinguish these assignments, starting with the $B\to S$
form factors   collected in Tab.~\ref{form}. For example in scenario
I, if we use the mixing coefficients in Eq.~(\ref{2}), the
production rates of $f_0(1710)$ and $f_0(1500)$ in $B$ decays are
much smaller than that of $f_0(1370)$ but they have large and
comparable production rates in $B_s$ decays; if we use the mixing
coefficients in Eq.~(\ref{3}), $f_0(1710)$ has small production
rates in both $B$ and $B_s$ decays but the other two mesons have
large and comparable production rates in $B$ and $B_s$ decays. Based
on our predictions on form
factors~\cite{Wang:2006ria,Li:2008tk,first}, these differences in
$B$ and $B_s$ decays are helpful to distinguish the two mixing
matrix.

\section{Glueball production in $B_c$ decays}

The ordinary light scalar meson is isospin singlet and/or flavor
SU(3) singlet, while the glueball is flavor SU(6) singlet. Therefore
it is difficult to distinguish them by the light $u$, $d$ and $s$
quark coupling. However, the light ordinary scalar meson has
negligible $c\bar c$ component, while the glueball have the same
coupling to $c\bar c$ as that to the  $u \bar u$, $d \bar d$ or $s
\bar s$. A clean way to identify a glueball is then through the
$c\bar c$ coupling to the glueball.

In $B$ decays, the initial heavy meson contains a light quark, thus
contributions of the gluon component always accompany with the quark
content $\bar nn$ or $\bar ss$. It is not easy to isolate the gluon
content. The situation in the doubly-heavy $B_c$ meson is different:
it contains a heavy charm antiquark. The semileptonic $B_c\to
f_0l\bar\nu$ decays would happen only through
Fig.~\ref{diagram:FeynBtof0}(a)(b) and (c) but not through
Fig.~\ref{diagram:FeynBtof0}(d) and (e). The observation of this
decay channel in the experiments will surely establish the existence
of a scalar glueball.   Moreover the CKM matrix element in this
channel is $V_{cb}$, thus the $B_c\to f_0l\bar\nu$ will have a
sizable branching ratio. This channel will depend on the $B_c\to G$
transition form factor which requires the less-constrained $B_c$
meson's light-cone distribution amplitude. But even if we assume the
form factor of $B_c\to G$ is smaller than the $B_c\to \eta_c$ form
factor by one order, branching ratios of $B_c\to G l\bar\nu
(l\bar\nu)$ are suppressed by two orders
\begin{eqnarray}
 {\cal B}(B_c\to Gl\bar\nu)& \sim & 1\% \times 0.01 =   10^{-4},
\end{eqnarray}
where the branching ratio of $B_c\to \eta_c l\bar\nu$  has been
taken as $1\%$. This branching ratio is large enough for the
experiments. One only needs to reconstruct the $f_0$ scalar meson in
the final state and also the $B_c$ meson mass in the intermediate
state, so that to make sure that the  scalar meson is produced from
two gluons. That experiment is achievable even if the $f_0$ meson is
not a pure glueball, but at least has a large portion of it.

$B_c\to f_0\pi^-$ is another potential mode to figure out the gluon
content. But in this mode, the $\bar nn$ component also contributes
through the annihilation diagrams. The $b$ and $\bar c$ quark
annihilates and the $d$ and $\bar u$ quark are created. The CKM
matrix element $V_{cb}$ and the Wilson coefficient $a_1$ are the
same with the emission diagram for the $B_c$-to-glueball transition.
The offshellnes of the two internal particles in annihilation
diagrams are of the order $m_{B_c}^2$. The electroweak vertex is the
$V-A$ type and the decay amplitude is proportional to the light
quark mass. Thus the decay amplitudes via annihilation diagram for
the $\bar nn$ component are expected to be suppressed.  As a result,
the $B_c\to f_0\pi^-$ also filters out the gluon component of the
scalar meson as an approximation.

\section{Summary}

Although the $B$-to-glueball form factors are small, they can not be
neglected and more interestingly these form factors may have
different interferences with those for the quark content, according
to different descriptions of scalar mesons. If a scalar meson is a
mixture of a glueball and an ordinary meson, we investigate the
possibility to extract the mixing mechanism from semileptonic $B$
decays. Semileptonic $B\to f_0l\bar \nu$ and $B_s\to f_0l^+l^-$
decays can be used to determine the internal structures. The
nonleptonic $B\to J/\psi f_0$ and $B_s\to J/\psi f_0$ decays are
also analyzed. To avoid the interference between the quark and the
gluon component, we find that the $B_c \to f_0 l\bar\nu$ and $B_c\to
f_0\pi^-$ will project out the gluon component of a scalar meson
cleanly. Our results can be generalized to the other glueballs.

\section*{Acknowledgement}

This work is partly supported by National Natural Science Foundation
of China under the Grant No. 10735080, and 10625525.

\end{document}